\documentclass[journal]{IEEEtran}
 
\usepackage{graphicx}
\usepackage{booktabs}
\usepackage{subfigure}
\usepackage{overpic}
\usepackage{amsmath}
\usepackage{flushend}
\usepackage{amssymb}
\usepackage{multirow}
\usepackage{makecell}
\usepackage{color, soul}
\soulregister\cite7

\usepackage[ruled]{algorithm2e}
\usepackage{hyperref}
\hypersetup{
	colorlinks=true,
	linkcolor=cyan,
	urlcolor=blue,
	citecolor=cyan,
}

\usepackage{colortbl}
\usepackage[dvipsnames]{xcolor}
\definecolor{Bg}{HTML}{e0f1ff}

\begin{document}

\title{Synthetic Aperture Radar Image Change Detection via Layer Attention-Based Noise-Tolerant Network}
\author{Desen Meng, Feng Gao, Junyu Dong, Qian Du, and Heng-Chao Li
\thanks{This work was supported in part by the National Key Research and Development Program of China under Grant 2018AAA0100602, and in part by the Natural Science Foundation of Shandong Province under Grant ZR2019QD011.

Desen Meng, Feng Gao, and Junyu Dong are with the School of Information Science and Engineering, Ocean University of China, Qingdao 266100, China. \emph{(Corresponding author: Feng Gao.)}

Qian Du is with the Department of Electrical and Computer Engineering, Mississippi State University, Starkville, MS 39762 USA.

H. -C. Li is with the Sichuan Provincial Key Laboratory of Information Coding and Transmission, Southwest Jiaotong University, Chengdu 610031, China.}}

\markboth{}%
{Shell }

\maketitle

\begin{abstract}

Recently, change detection methods for synthetic aperture radar (SAR) images based on convolutional neural networks (CNN) have gained increasing research attention. However,  existing CNN-based methods neglect the interactions among multilayer convolutions, and errors involved in the preclassification restrict the network optimization. To this end, we proposed a layer attention-based noise-tolerant network, termed LANTNet. In particular, we design a layer attention module that adaptively weights the feature of different convolution layers. In addition, we design a noise-tolerant loss function that effectively suppresses the impact of noisy labels. Therefore, the model is insensitive to noisy labels in the preclassification results. The experimental results on three SAR datasets show that the proposed LANTNet performs better compared to several state-of-the-art methods. The source codes are available at \url{https://github.com/summitgao/LANTNet}.

\end{abstract}


\IEEEpeerreviewmaketitle

\section{Introduction}

\IEEEPARstart{R}{emote} sensing image change detection aims to identify regions that changed in the same area using multitemporal images.  In fact, it facilitates various applications, such as disaster assessment, deforestation monitoring, and object detection \cite{lv22grsl} \cite{wang22jstars} \cite{ai19tgrs}. The synthetic aperture radar (SAR) is an ideal source for change detection, since the SAR sensor is independent of sunlight illumination and weather conditions. Therefore, SAR change detection has received extensive research attention recently. 

\begin{figure}
  \centering
  \includegraphics[width=3.3in]{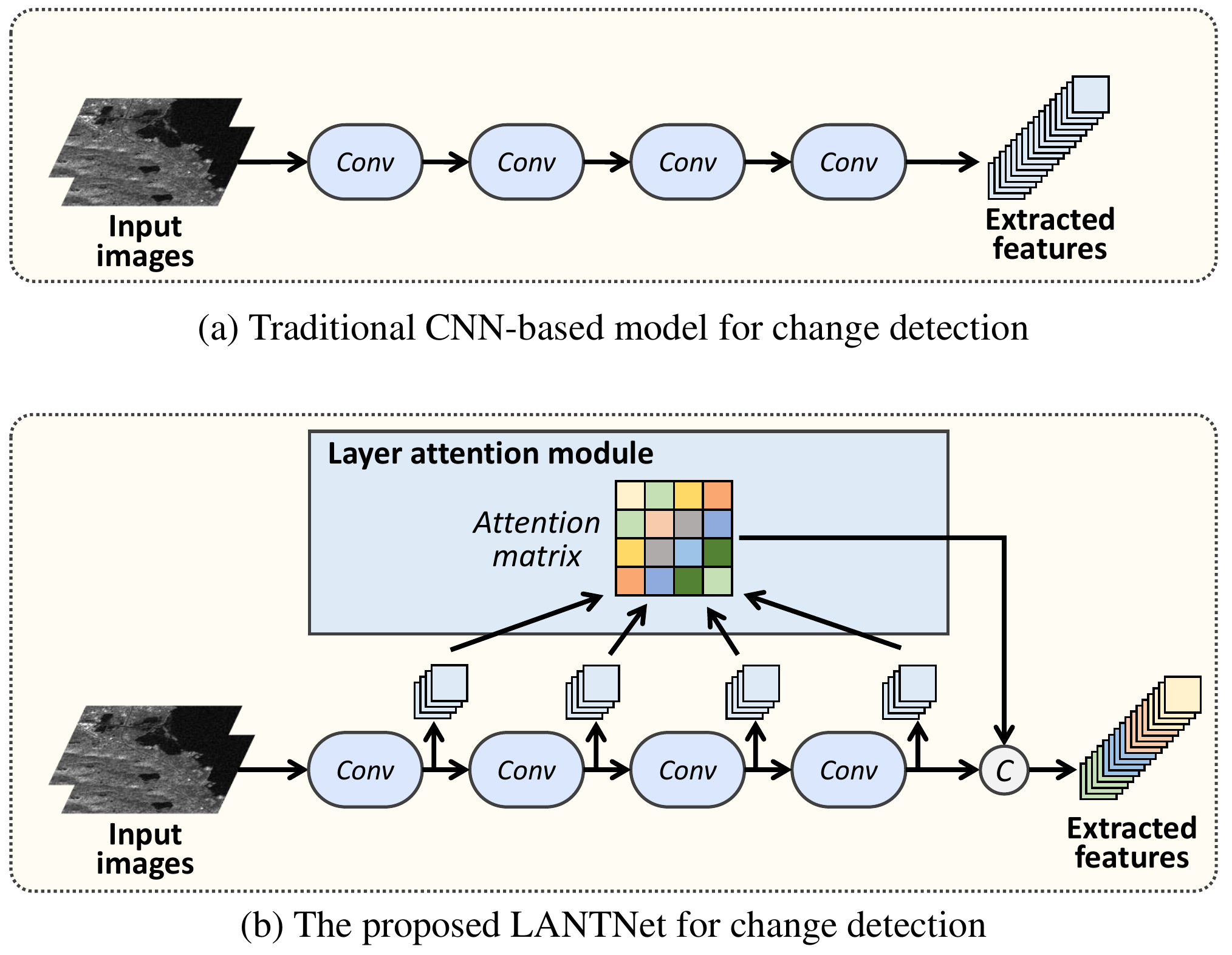}
  \caption{Comparisons of the traditional method with the proposed LANTNet. (a) The traditional CNN-based feature extraction model for change detection. They overlook the interactions among multi-layer convolutions. (b) The proposed LANTNet. The layer attention module adaptively computes weights for features from different layers and thus improves the feature representations.}
  \label{fig_comp}
\end{figure}

Many supervised and unsupervised methods have been devoted to detect changes in multitemporal SAR images. In fact, it commonly costs a lot of effort to collect high-quality labeled samples that reflect the changed information, and learning the changed information in an unsupervised manner is more efficient. Therefore, in this letter, we focus on unsupervised SAR change detection. 

Traditionally, the task of SAR image change detection are solved by clustering methods, such as $k$-means \cite{5196726}, fuzzy $c$-means (FCM) \cite{gong14tfs} and hierarchical clustering \cite{gao16grsl}. Recently, leveraging the representation power of convolutional neural networks (CNNs) and the attention mechanism to capture changed information has become a common solution for change detection. Gao et al. \cite{gao21grsl} designed an adaptive fusion convolution module to encode the input, allowing the network to focus on the more informative channels. Zhang et al \cite{zhang22sparse} proposed a sparse feature clustering network for SAR image change detection. Geng et al. \cite{geng19saliency} presented a saliency-guided network to suppress the influence of speckle noise and improve the SAR change detection performance. In \cite{gao19transfer}, transfer learning is employed for the parameter initialization for change detection deep networks. Furthermore, wavelet-CNN \cite{8641484}, discrete cosine transform \cite{9420150}, and ensemble learning \cite{wang21ensemble} are also used to enhance the representation power of deep networks.

\begin{figure*}[ht]
\centering
\includegraphics [width=5in]{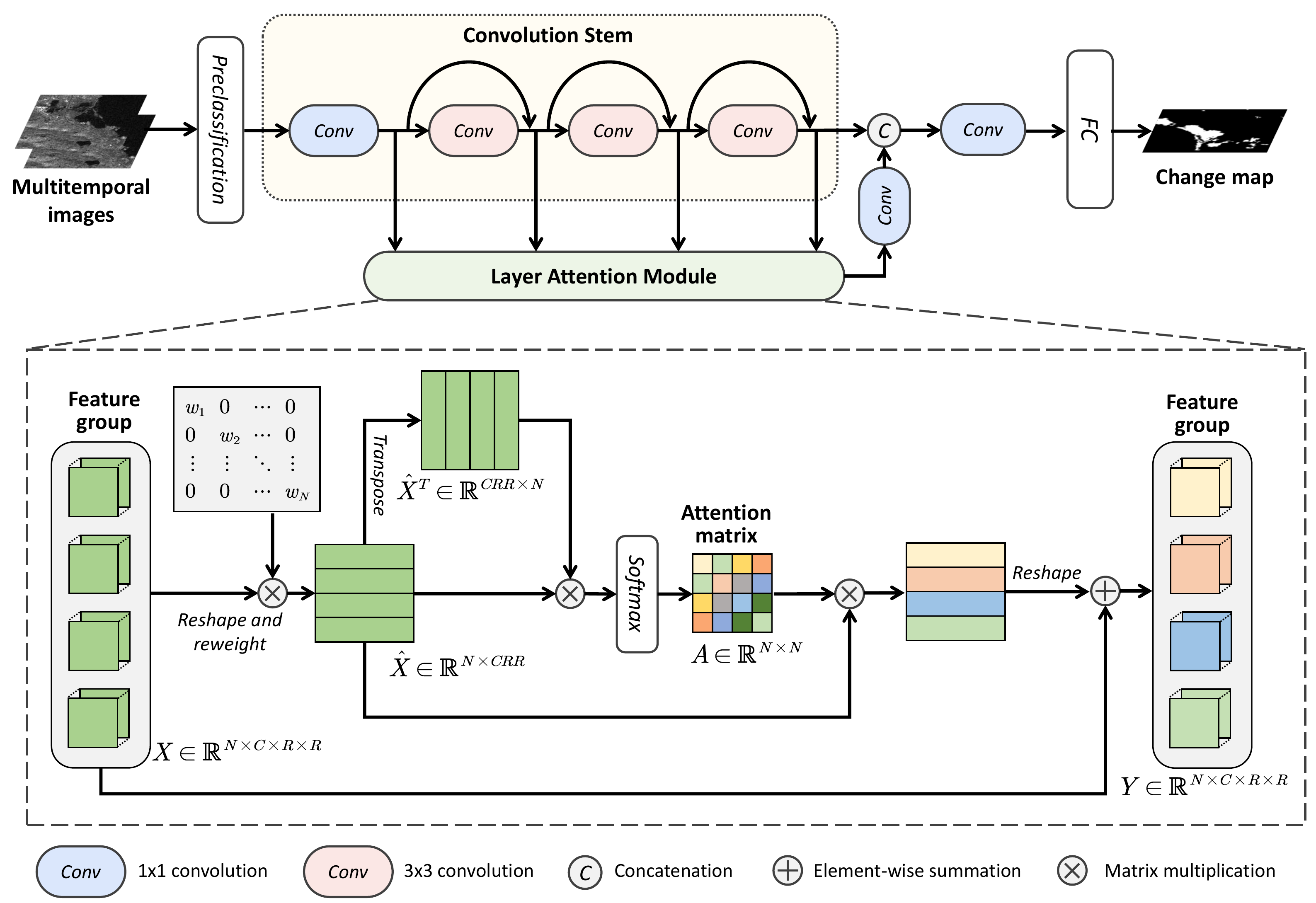}
\caption{The pipeline of LANTNet. It is comprised of the preclassification, convolution stem, layer attention module and the FC layer. The convolution stem consists of four convolutional layers. Correspondingly, the layer attention module uses four feature groups to exploit the correlations among the four-layer features generated by the convolution stem.} 
\label{fig_pipeline}
\end{figure*}

Indeed, the CNN-based models have shown impressive performance in change detection task. However, they still show limitations in the following aspects: \textbf{1) Feature interaction among convolution layers.} Research in human vision system finds that visual representations in the cortex are activated in parallel fashion \cite{chelazzi98}. Existing self-attention-enhanced CNNs commonly incorporate an attention block after a convolution layer, which overlooks the interactions among multi-layer convolutions. In light of this, it is essential to explore the information flow among the convolutional layers. \textbf{2) Errors involved in the pseudo-labels.} Existing unsupervised change detection methods generally use clustering to generate pseudo-labels with high confidence for training. However, the errors involved in the pseudo-labels restrict network optimization \cite{zhang2018generalizedGCE}. Therefore, it is vital to build a noise-tolerant model for unsupervised change detection task.

To address the aforementioned issues, we propose a \textbf{L}ayer \textbf{A}ttention-based \textbf{N}oise-\textbf{T}olerant \textbf{Net}work, dubbed as LANTNet, which interconnects convolutional layers and is less sensitive to noisy labels. Specifically, we design a layer attention module for change detection, which adaptively weights features from different layers, as illustrated in Fig. \ref{fig_comp}. In addition, we design a noise-tolerant loss function that can effectively suppress the impact of noisy labels. It combines the Cross Entropy (CE) loss and the Mean Absolute Error (MAE). Hence, the network is less sensitive to noisy labels and has a fast convergence rate. We conducted extensive experiments on three SAR datasets, and the results demonstrate that the proposed LANTNet yields better performance as compared to state-of-the-art methods. 

In summary, the contributions of this letter are threefold: 

\begin{itemize}
	
\item We propose a layer attention module, which exploits the correlations of multi-layer convolutions. Features from different convolution layers cooperate with each other, effectively improving the representation power of the network.

\item We introduce a noise-tolerant loss function to alleviate the influence of noisy labels in pseudo-labeled samples. It makes the model less sensitive to noisy labels and has fast convergence rate. 

\item Experimental results confirm the effectiveness of the proposed LANTNet. Furthermore, we have made our data and codes publicly available to benefit other research of the remote sensing community.

\end{itemize}

\section{Methodology}

Given the multitemporal SAR images $I_1$ and $I_2$ acquired in the same geographical zone, the goal of change detection is to generate a change map where each spatial location is assigned to one type of change. In this work, we consider the generation of change map as a binary classification task, which means that the label is either 1 (changed) or 0 (not changed).

As shown in Fig. \ref{fig_pipeline}, the proposed LANTNet comprises three components: preclassification module, convolutional stem and layer attention module. First, multitemporal coregistered SAR images $I_1$ and $I_2$ are fed into the preclassification module to obtain samples that are highly likely to be changed or unchanged. New image patches will be generated and then fed into the convolutional stem to compute feature groups $X$. The feature groups are then updated to attention features $Y$ by the layer attention module. Finally, $Y$ and the features of the convolutional stem are combined for classification to calculate the final change map.

\subsection{Preclassification and Convolutional Stem}

In the preclassification module, the logarithmic ratio operator and the hierarchical Fuzzy C-Means Algorithm are employed, and details of the module can be found in \cite{gao16grsl}. After preclassification, the image patches ($R \times R$ pixels) with pseudo-labels are extracted from $I_1$, $I_2$, and the difference image, respectively. They are concatenated to form new image patches ($3 \times R \times R$ pixels) as training samples.

The convolutional stem consists of four convolution layers. 1$\times$1 convolution is first used to extract the shallow feature $F_0$. Then, three 3$\times$3 convolution layers are employed to extract intermediate features $F_i$ as
\begin{equation}
    F_i=\textrm{Conv}(F_{i-1}), ~~~ i= 1, 2, 3.
\end{equation}
For the shallow feature $F_0$, we set the channel numbers to 16, and for intermediate features $F_1$, $F_2$, and $F_3$, we set the channel numbers to 32. Then, $1\times1$ convolution is employed to increase the channel dimensions of $F_0$ to 32. After that, $F_0$, $F_1$, $F_2$, and $F_3$ are combined to form a feature group, and fed into the layer attention module.

\subsection{Layer Attention Module}

The attention mechanism helps the network focus on the most important part and neglect irrelevant information. However, the correlations among multi-layer convolutions are rarely taken into account. To this end, we design a layer attention module to weight the features $F_0$, $F_1$, $F_2$, and $F_3$ from multiple layers by learning the correlation matrix. Therefore, layer attention effectively fuses the spatial information of low-level layers and the semantic information from high-level layers. The informative layers are emphasized, and the redundant ones are suppressed. Fig. \ref{fig_pipeline} shows the details of the proposed layer attention module. The input of the module is the feature group $X\in \mathbb{R}^{N\times C\times R\times R}$ from the convolution stem. In our implementation, $N$ is set to 4. Then we reshape $X$ to a feature matrix with the dimension of $N\times CRR$. Subsequently, we apply matrix multiplication between the new feature matrix and a diagonal matrix $W$ to weight the features from different layers. Note that $W$ is initialized to identity matrix and only the diagonal elements of $W$ are updated in each epoch. The weight matrix adaptively assign weights to the input feature groups. After that, a weighted feature matrix $\hat{X}\in\mathbb{R}^{N\times CRR}$ can be obtained. 

Next, we calculate an attention matrix $A\in \mathbb{R}^{N\times N}$ as the correlation matrix as follows:
\begin{equation}
    A = \textrm{Softmax}(\hat{X}\otimes \hat{X}^{T}),
\end{equation}
where the element $A_{i,j}$ in the attention matrix denotes the correlation index between the $i$-th and $j$-th layer features, and $\otimes$ denotes the matrix multiplication. Matrix multiplication is performed between the weighted feature matrix $\hat{X}$ and its transpose matrix $\hat{X}^T$. Softmax operation is applied to generate the attention matrix $A$.

Finally, we multiply the weighted feature matrix $\hat{X}$ by the attention matrix $A$, and reshape the output to the dimension of $N\times C \times R\times R$. The input $X$ are combined to form the final output $Y$ as:
\begin{equation}
    Y = \phi(A\otimes \hat{X})\oplus X,
\end{equation}
where $\phi$ denotes the reshape operation.

To compute the final change map, $Y$ is reshaped to the dimension of $NC\times R \times R$, and then fed into a $1\times1$ convolution layer for channel reduction, as illustrated in Fig. \ref{fig_pipeline}. After that, the combined features are handled by a fully connected (FC) layer. The softmax function is used to calculate the possibility of changed or unchanged. After training, the final change map can be obtained by classifying all pixels from the multitemporal SAR images.

\subsection{Noise-tolerant loss function}

Like most existing unsupervised methods, we employ clustering to generate pseudo-labels with high probability for network training. However, errors are commonly involved in these pesudo-labels, and restrict the network optimization \cite{zhang2018generalizedGCE}. Therefore, it is critical to build a noise-tolerant model.

Toward this end, we design a noise-tolerant loss which combines the Cross Entropy (CE) loss and the Mean Absolute Error (MAE). Suppose the dataset for change detection is represensted as $\{ (x_i, y_i)|1\leq i\leq m \}$, where $x_i$ is the input image patch, $y_i$ is the label associated with $x_i$, and $m$ is the total number of samples. The MAE loss is defined as:
\begin{equation}
    \mathcal{L}_\textrm{MAE}(f(x),y)= \|e_y-f(x)\|_1,
\end{equation}
where $f(\cdot)$ represents the network and $e_y$ is a one-hot vector with $e_{yj}=1$ if $j=y$, otherwise $e_{yj}=0$.

The CE loss is computed as:
\begin{equation}
    \mathcal{L}_\textrm{CE}(f(x),y)=-e_y\log f(x).
\end{equation}

Ghosh et al. \cite{ghosh2017robust} empirically demonstrated that MAE is robust to noisy labels, but it often costs more time to converge. To the contrary, CE tends to overfit noisy labels, but it performs better than MAE when training with clean data. To leverage the good convergence performance of CE, and exploit the noise robustness of provided by MAE, we design a robust loss function which combines the CE and MAE. Formally, the robust loss function is defined as:
\begin{equation}
    \mathcal{L}=\alpha \mathcal{L}_\textrm{CE}+\beta \mathcal{L}_\textrm{MAE}
\end{equation}
where $\alpha$ and $\beta$ are two decoupled hyperparameters.  In our implementation, we set $\alpha$ to 0.1 and $\beta$ to 0.9 for a trade-off between convergence efficiency and the noise robustness. 

\section{Experiments and Analysis}

\subsection{Datasets and Evaluation Metrics}

To evaluate the proposed LANTNet, we conducted extensive experiments on three real SAR datasets captured by different sensors. The first dataset is Chao Lake dataset with the size of $384\times 384$ pixels. It was captured by the Sentinel-1 sensor in May and July 2020 in China, respectively. The second dataset is the Ottawa dataset with the size of $290\times 350$ pixels, which was acquired by the Radarsat sensor in May and August 1997, respectively. The Ottawa dataset shows the changed areas afflicted by floods. The third dataset is the Sulzberger dataset which was captured by the Envisat satellite of the European Space Agency on March 11 and 16, 2011. Both images show a breakup of the ice shelf caused by a tsunami.

Five commonly-used evaluation metrics, i.e., false positive (FP), false negative (FN), overall error (OE), percentage of correct classification (PCC) and Kappa coefficient (KC) are used to evaluate the change detection performance.

\subsection{Analysis of the Patch Size}

\begin{figure}
  \centering
  \includegraphics[width=2.5in]{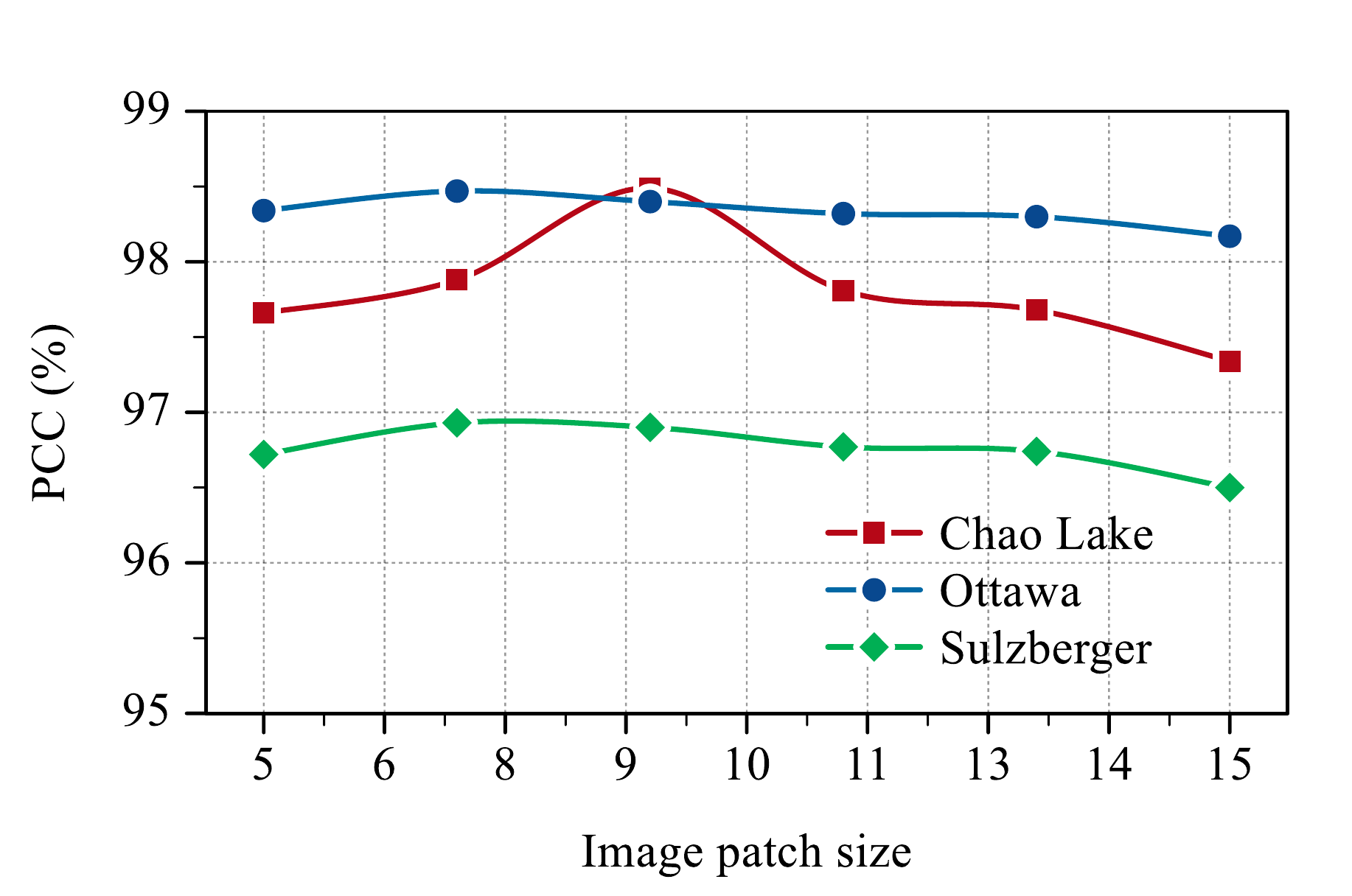}
  \caption{Relationship between image patch size and the PCC value.}
  \label{fig_patchsize}
\end{figure}

The image patch size $R$ is a critical parameter. To explore the relationship between the patch size $R$ and PCC, we set $R$ from 5 to 15, and the experimental results are shown in Fig. \ref{fig_patchsize}. It can be seen that all curves tend to increase first and then decrease. When $R$ is small, the contextual information around each pixel is insufficient. However, if $R$ becomes too large, irrelevant neighborhood information would be taken into account. Therefore, we set $R=9$ for the Chao Lake dataset, and $R=7$ for the Ottawa and Sulzberger datasets.

\begin{figure}
  \centering
  \includegraphics[width= 2.8in]{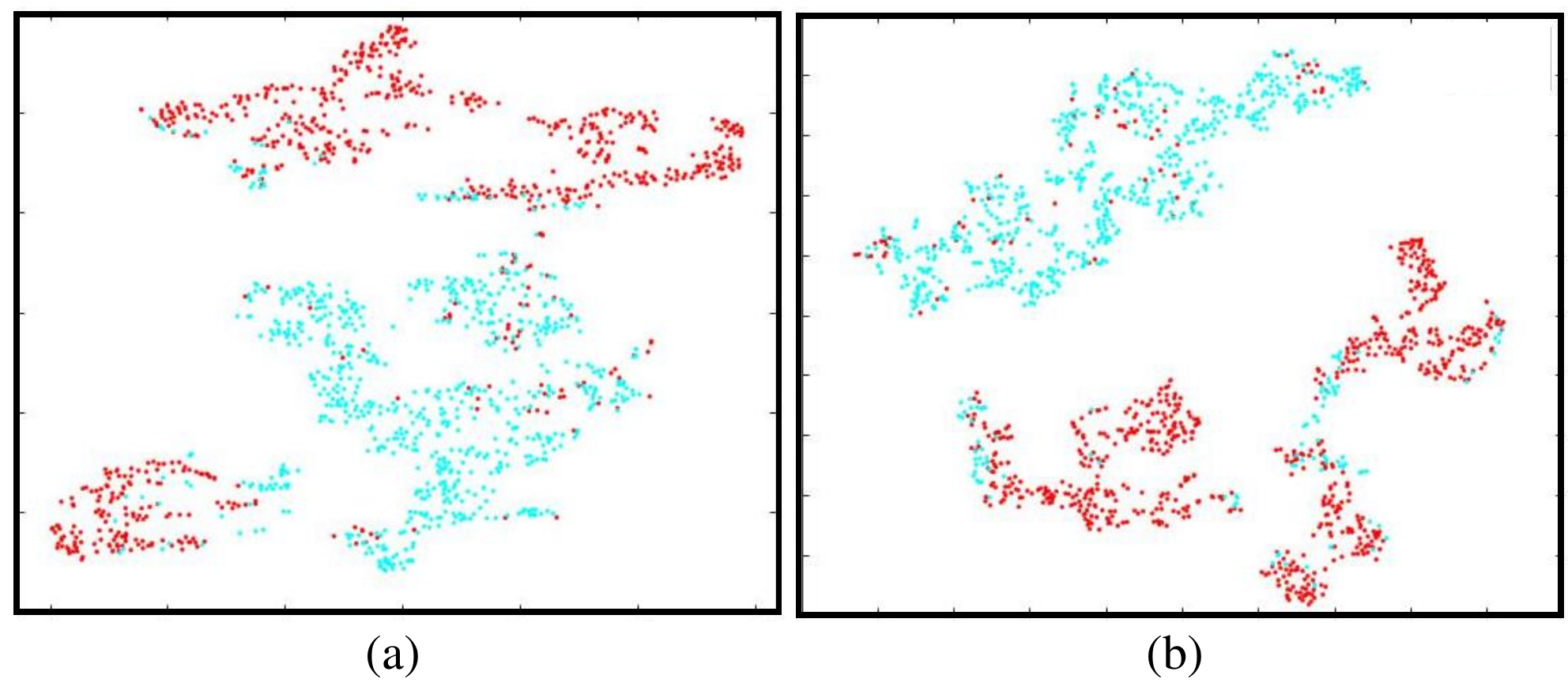}
  \caption{Visualization of the feature representations on the Ottawa dataset. (a) Features before the layer attention module. (b) Features after the layer attention module.}
  \label{fig_visual}
\end{figure}

\begin{table}[h]
\centering
\caption{Ablation Studies of the Proposed LANTNet}
\label{table_ablation}
\begin{tabular}{c|c c c} 
\toprule
\multirow{2}{*}{Method} 
    & \multicolumn{3}{c}{PCC on different datasets ($\%$)} \\ \cmidrule{2-4}
& Chao Lake & Ottawa & Sulzberger\\ 
\midrule
Basic CNN    & 97.49 & 98.22  &96.36   \\
w/o layer attention   & 97.82  & 98.32  & 96.64    \\  
w/o noise-tolerant loss  & 97.98  & 98.42  & 96.69    \\ 
\rowcolor{Bg} Proposed LANTNet & 98.49  & 98.47  & 96.93    \\  
\bottomrule
\end{tabular}
\end{table}

\begin{figure*}[htb]
  \centering
  \includegraphics[width=5.5in]{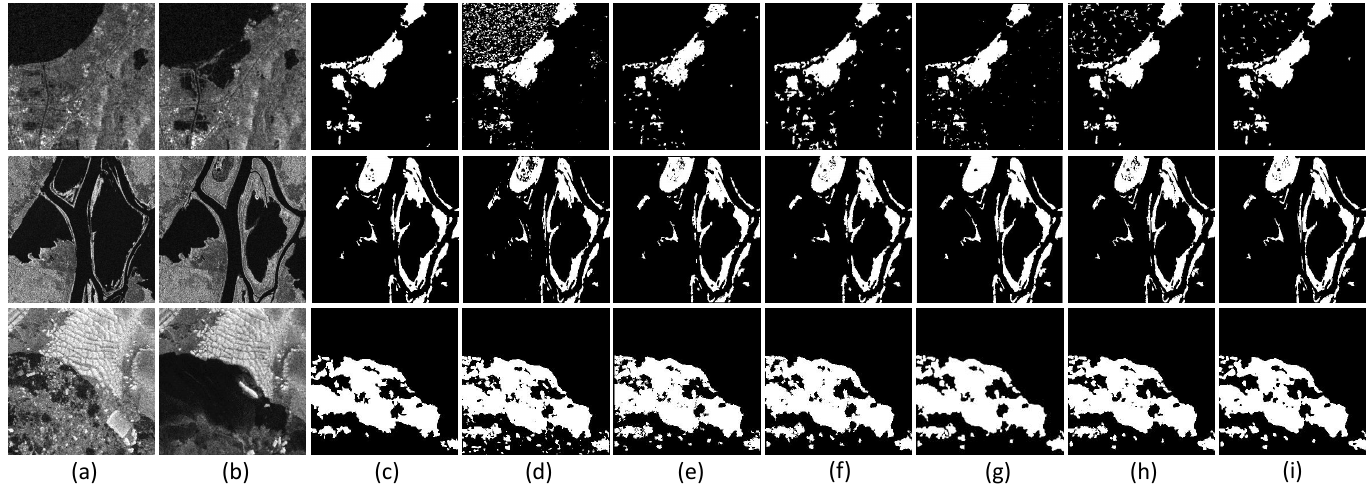}
  \caption{Visualized results of different change detection methods on the Chao Lake dataset (first row), Ottawa dataset (second row) and Sulzberger dataset (third row) :  (a) Image captured at $t_1$. (b) Image captured at $t_2$. (c) Ground truth image. (d) Result by PCAKM \cite{5196726}. (e) Result by NR-ELM \cite{gao16jars}. (f) Result by GaborPCANet \cite{gao16grsl}. (g) Result by CWNN \cite{8641484}. (h) Result by DDNet \cite{9420150}. (i) Result by the proposed LANTNet.}
  \label{fig_result}
\end{figure*}

\subsection{Ablation Study}

To verify the effectiveness of the proposed layer attention module and the noise-tolerant loss, we conduct ablation experiments on three datasets. Three variants are designed for comparison. First, we design a basic CNN with the same structure as LANTNet, which removes the layer attention module and uses cross entropy as loss function instead of the noise-tolerant loss. Then we run our model without layer attention module (\emph{w/o layer attention}) and without the noise-tolerant loss(\emph{w/o noise-tolerant loss}).  The results are shown in Table \ref{table_ablation}. It can be observed that both the layer attention module and the noise-tolerant loss are effective for change detection performance improvement. 

Furthermore, to gain deeper insights toward the layer attention, we visualized the features before and after the layer attention module with the tool of t-SNE. As shown in Fig. \ref{fig_visual}, the distance between features clusters derived by the layer attention is larger, which reflects the feature representations by our model are indeed more discriminative.

\begin{table}[h]
    \centering
	\caption{Change Detection Results of Different Methods on Three Datasets}
	\label{table_res}
    \begin{tabular}{c|c c c c c} 
    \toprule
    \multirow{2}{*}{Method} 
    & \multicolumn{5}{c}{Results on the Chao Lake dataset}\\
    \cmidrule{2-6}
      & FP & FN & OE & PCC  & KC\\ 
    \midrule
     PCAKM \cite{5196726}   & 8521 & 2248 & 10769 & 92.70 & 65.58 \\
     NR-ELM \cite{gao16jars}   & 595 & 3836 & 4431 & 97.00 & 81.27 \\
     GaborPCANet \cite{gao16grsl}    & 2946 & 1771 & 4717 & 96.80 & 82.66 \\
     CWNN \cite{8641484}   & 959 & 2397 & 3356 & 97.72 & 86.63\\
     DDNet \cite{9420150}   & 2097 & 1068 & 3165 & 97.85 & 88.31 \\
     \rowcolor{Bg} Proposed LANTNet  & 1358 & 867 & 2225 & 98.49 & 91.65\\
    \bottomrule
    \toprule
    \multirow{2}{*}{Method} & \multicolumn{5}{c}{Results on the Ottawa dataset} \\
    \cmidrule{2-6}
      & FP & FN & OE & PCC  & KC\\ 
    \midrule
     PCAKM \cite{5196726}  & 729 & 2458 & 3187 & 96.86 & 87.67  \\
     NR-ELM \cite{gao16jars}   & 517 & 1310 & 1827 & 98.20 & 93.10  \\
     GaborPCANet \cite{gao16grsl} & 755 & 1080 & 1835 & 98.19 & 93.15 \\
     CWNN \cite{8641484} & 1291 & 434 & 1725 & 98.30 & 93.75 \\
     DDNet \cite{9420150} & 504 & 1191 & 1695 & 98.33 & 93.62 \\
     \rowcolor{Bg} Proposed LANTNet  & 638 & 911 & 1549 & 98.47 & 94.23 \\
    \bottomrule
    \toprule
    \multirow{2}{*}{Method} 
    & \multicolumn{5}{c}{Results on the Sulzberger dataset} \\ 
    \cmidrule{2-6}
      & FP & FN & OE & PCC  & KC\\ 
    \midrule
     PCAKM \cite{5196726} & 3308 & 701 & 4009 & 93.88 & 84.49\\
     NR-ELM \cite{gao16jars} & 2386 & 646 & 3032 & 95.37 & 88.07\\
     GaborPCANet \cite{gao16grsl} & 2485 & 494 & 2979 & 95.45 & 88.34\\
     CWNN \cite{8641484} & 1598 & 1132 & 2730 & 95.83 & 88.98\\
     DDNet \cite{9420150} & 1710 & 526 & 2236 & 96.59 & 91.10\\
     \rowcolor{Bg} Proposed LANTNet & 1229 & 784 & 2013 & 96.93 & 91.87\\
    \bottomrule
    \end{tabular}
\end{table} 

\subsection{Experimental Results and Comparison}

To verify the effectiveness of the proposed LANTNet, five closely related change detection methods are used for comparison, including PCAKM \cite{5196726}, NR-ELM \cite{gao16jars}, GaborPCANet \cite{gao16grsl},  CWNN \cite{8641484}, and DDNet \cite{9420150}. Note that the methods mentioned above are implemented with the default parameters provided in their work. All visual results are illustrated in Fig. \ref{fig_result}, and the corresponding quantitative results are shown in Table \ref{table_res}.

On the Chao Lake dataset (the first row of Fig. \ref{fig_result}),  PCAKM has the highest FP value among all methods and the FP value affects the overall performance of PCAKM. At the same time, GaborPCANet and DDNet also suffer from high FP values, while NR-ELM and CWNN suffer from high FN values. CWNN and DDNet are two CNN-based methods and have strong feature representation power, as shown in Table \ref{table_res}. However, the pseudo-labels utilized by CWNN and DDNet are not accurate enough, and their models may overfit the noisy labels. The KC value of the proposed LANTNet is impoved by 26.07$\%$, 10.38$\%$, 8.99$\%$, 5.02$\%$ and 3.34$\%$ over PCAKM, NR-ELM, GaborPCANet, CWNN and DDNet, respectively. It is evident that the proposed method achieves the best performance on the Chao Lake dataset.

On the Ottawa dataset (the second row of Fig. \ref{fig_result}), PCAKM, NR-ELM, GaborPCANet, and DDNet suffer from high FN values and many change pixels are not correctly classified. CWNN suffers from a high FP value and generates many noisy spots in the change map. The proposed LANTNet achieves the best PCC value, and it demonstrates that layer attention effectively improves the feature representation. The comparisons demonstrate the good performance of the proposed LANTNet on the Ottawa dataset.

On the Sulzberger dataset (the third row of Fig. \ref{fig_result}), all the methods suffer from high FP values, and the proposed LANTNet achieves the lowest FP value. It can be seen from Fig. \ref{fig_result} that the proposed LANTNet is effective in suppressing the label noise and reduces the FP values. The KC value of the proposed LANTNet is impoved by 7.38$\%$, 3.8$\%$, 3.53$\%$, 2.89$\%$ and 0.77$\%$ over PCAKM, NR-ELM, GaborPCANet, CWNN and DDNet, respectively. The comparison shows that the LANTNet is powerful in noise suppression on the Sulzberger dataset.

\section{Conclusions}

In this letter, we propose a layer attention-based noise-tolerent neural network (LANTNet) for SAR change detection. Compared with existing CNN-based methods, our LANTNet improves in two aspects. First, we design a layer attention module that adaptively weights the feature of different convolution layers. In addition, since the pseudo-labels obtained by preclassification are not accurate enough, we design a noise-tolerant loss to help the network avoid overfitting to the noisy labels. Experimental results on three multi-temporal SAR datasets demonstrate that the proposed LANTNet is superior to several state-of-the-art change detection methods. 

\bibliography{re} 
\bibliographystyle{IEEEtran}

\end{document}